\documentclass[aps,prl,twocolumn,groupedaddress]{revtex4-1}
\usepackage{graphicx,epsfig,dcolumn,afterpage}
\usepackage{colortbl}
\usepackage[latin1]{inputenc} 
\usepackage{amsmath}
\usepackage{amssymb}  %gtrsim

\usepackage{color}
\usepackage{colortbl}

\newcommand{\TV}[1]{}  %if something shall be left out

\newcommand{\xten}[1]{{}\times 10^{#1}}
\newcommand{\csixty}[0]{C$_{60}$}
\newcommand{\Necsixty}[0]{Ne@C$_{60}$}

\newcommand{\eV}{\,\text{eV}}

\newcommand{\au}{\,\text{a.u.}}

\newcommand{\Wcm}{\,\text{W}/\text{cm}^2}
\newcommand{\as}{\,\text{as}}
\newcommand{\fs}{\,\text{fs}}

\begin{document}

\title{Observation of molecular dipole excitations by attosecond self-streaking}

\author{Georg Wachter$^1$}
\email{georg.wachter@tuwien.ac.at}
\author{Stefan Nagele$^1$}
\author{Shunsuke A. Sato$^2$}
\author{Renate Pazourek$^4$}
\author{Michael Wais$^1$}
\author{Christoph Lemell$^1$}
\author{Xiao-Min Tong$^{2,3}$}
\author{Kazuhiro Yabana$^{2,3}$}\
\author{Joachim Burgd\"orfer$^{1,5}$}

\affiliation{$^1$Institute for Theoretical Physics, Vienna University of Technology, 1040 Vienna, Austria, EU}
\affiliation{$^2$Graduate School of Pure and Applied Sciences, University of Tsukuba, Tsukuba 305-8571, Japan}
%\affiliation{$^3$Institute of Materials Science, Graduate School of Pure and Applied Sciences, University of Tsukuba, Tsukuba 305-8573, Japan}
\affiliation{$^3$Center for Computational Sciences, University of Tsukuba, Tsukuba 305-8577, Japan}
\affiliation{$^4$Department of Physics and Astronomy, Louisiana State University, Baton Rouge, Lousiana, USA}
\affiliation{$^5$Institute of Nuclear Research of the Hungarian Academy of Sciences (ATOMKI), Debrecen H-4001, Hungary, EU}

\date{\today}

\begin{abstract}
We propose a protocol to probe the ultrafast evolution and dephasing of coherent electronic excitation in molecules in the time domain by the intrinsic streaking field generated by the molecule itself.
Coherent electronic motion in the endohedral fullerene \Necsixty~is initiated by a moderately intense femtosecond UV-VIS pulse leading to coherent oscillations of the molecular dipole moment that persist after the end of the laser pulse. 
The resulting time-dependent molecular near-field is probed through the momentum modulation of photoemission from the central neon atom by a time-delayed attosecond XUV pulse.
Our ab-initio time-dependent density functional theory and classical trajectory simulations predict that this self-streaking signal accurately traces the molecular dipole oscillations in real time. We discuss the underlying processes and give an analytical model that captures the essence of our ab-initio simulations. 
\end{abstract}

% c60 dipole streaking

% 32.80.fb - Atomic properties and interactions with photons  - Photoionization of atoms and ions
% 42.65.Re - optics - Ultrafast processes; optical pulse generation and pulse compression
% 32.80.Rm - Multiphoton ionization and excitation to highly excited states
% 33.80.Eh - Molecular properties and interactions with photons - Autoionization, photoionization, and photodetachment
% photogalvanic effect
%\pacs{42.50.Hz,42.65.Re,42.65.Pc,78.47.J-}
% 42.65.Re ultrafast processes in nonlinear optics
% 42.65.Pc switching in nonlinear optics
% 78.47.J-  ultrafast processes in solid-state dynamics
% 72.40.+w photovoltaic effect in bulk matter

% OLD \pacs{71.15.Mb, 42.50.Hz, 78.47.J-, 72.20.Ht}
% DFT, strong field, ultrafast in solids, conductivity in insulators / high field, nonlin effects.

\maketitle

In recent years, the availability of wave-form controlled few-cycle intense laser pulses has afforded novel opportunities to investigate the ultrafast and non-linear optical response of matter. Studies of electronic motion in the time domain have focused on rare-gas atoms \cite{krausz_attosecond_2009} and molecules \cite{scrinzi_attosecond_2006,lepine_attosecond_2014} and, more recently, on nanostructures, surfaces, and bulk matter \cite{gertsvolf_orientation-dependent_2008,zherebtsov_controlled_2011}. With the advent of attosecond XUV pulses synchronized to intense optical pulses, attosecond streaking \cite{hentschel_attosecond_2001,kienberger_atomic_2004} has yielded insight into the real-time motion of electrons during photoionization \cite{schultze_delay_2010}. % where the electron trajectory can be modified due to the atomic dipole moment of the residual ion \cite{pazourek_attosecond_2012}. 
Attosecond \emph{surface} streaking \cite{cavalieri_attosecond_2007,neppl_direct_2015} has found photoelectron spectra to be highly sensitive to the atomic-scale surface polarization and near-field distribution. These results nurture the hope that attosecond streaking can also extract information on induced dipoles and near-fields in nano-scale systems bridging the gap between single atoms and extended matter. 
One challenge for such proposed ``nano-plasmonic streaking'' \cite{skopalova_numerical_2011,sussmann_attosecond_2011} is the smearing out of the streaking signal over the nanometric target including emission from deeper layers. 
%To counteract this, recent surface streaking experiments \cite{neppl_direct_2015} have placed different atomic species at engineered positions in the sample, thereby correlating individual photoemission lines with the spatial position of the initial electron wave function. 

In this work, we analyze a scenario that allows to probe molecular coherent dipole excitation with sub-nanometer spatial resolution and attosecond time resolution by self-streaking. As prototypical case we study the endohedral fullerene \Necsixty \cite{saunders_stable_1993}. 
The \csixty~cage serves as a nano-sized system with strong near-fields for which an ab-initio treatment is still feasible. The central neon atom serves as a distinct atomic-scale localized source of electrons to be ionized by an attosecond XUV pulse that probe the motion of the cage electrons. While \emph{transient} molecular dipole moments \emph{during} the IR laser pulse have recently been studied by Neidel et al.~\cite{neidel_probing_2013}, we employ here UV-VIS radiation to access low-lying electronic excitations of the molecule. The resulting coherent wavepacket motion leads to an oscillating time-dependent dipole moment sustained after the conclusion of the UV-VIS laser pulse. 
This time-dependent dipole moment causes an appreciable intrinsic streaking field for photoelectrons emitted from the central atom by a delayed attosecond XUV pulse. 
The streaking signal is found to trace the instantaneous molecular dipole moment after the laser pulse, enabling the study of electronic excitations, their evolution, dephasing, and decay with sub-femtosecond time resolution. 

In standard attosecond streaking \cite{itatani_attosecond_2002}, the attosecond XUV pulse acts as ``pump'' which ionizes an electron whose emission time is probed by the simultaneously present near-IR probe pulse. In the present scenario, their roles are reversed: 
the UV-VIS pulse acts as a pump to excite the molecule, and the XUV pulse probes the molecular near-field associated with this excitation by a momentum shift of the outgoing electron (``self-streaking'') somewhat resembling attosecond transient absorption \citep{pfeifer_time-resolved_2008,goulielmakis_real-time_2010,wirth_synthesized_2011}. It can be viewed as a variant of time-resolved photoelectron spectroscopy \citep{stolow_femtosecond_2004,de_giovannini_simulating_2013} or as the generalization of the ``nanoplasmonic-field microscope'' \cite{stockman_attosecond_2007} from surface plasmons to the study of single molecules. 
The key advantage of self-streaking is that time-resolved probing of the coherent electron dynamics proceeds in the absence of any distorting external NIR field.
%A somewhat similar idea has recently been proposed for the study of the onset of strong-field excitation \cite{mignolet_control_2014}, however requiring an attosecond pulse train with tunable time delay between the attosecond pulses.

The electron dynamics in \Necsixty~driven by the UV-VIS field are treated within time-dependent density functional theory \cite{runge_density-functional_1984,parr_density-functional_1994,ullrich_time-dependent_2011,marques_fundamentals_2012}. The simulation has been described elsewhere \cite{yabana_time-dependent_1996,kawashita_oscillator_2009}. Briefly, we solve the time-dependent Kohn-Sham equations 
\begin{eqnarray} \label{eq:tdks}
i \partial_t \psi_i(\mathbf r,t) & = & \bigg\{ - \frac{1}{2} \nabla^2 + V_\mathrm{ion} + F_L(t) z + \nonumber \\ 
                                 &   & \int d\mathbf{r}'  \frac{n(\mathbf r', t)}{|\mathbf r - \mathbf r'|} + V_\mathrm{XC}[n(\mathbf r, t)] \bigg \} \psi_i(\mathbf r, t) \quad ,
\end{eqnarray}
where the terms on the right hand side correspond to the kinetic energy, the potential of the carbon and neon 1s$^2$ cores in terms of norm-conserving pseudopotentials \cite{troullier_efficient_1991}, the coupling to the UV-VIS laser field $F_L(t)$ in dipole approximation and length gauge, the Hartree potential, and the exchange-correlation (XC) potential for which we employ the adiabatic local density approximation \cite{perdew_self-interaction_1981}. The valence electron density is $n(\mathbf r, t) = \sum_i |\psi_i(\mathbf r, t)|^2$. 
Eqs.~(\ref{eq:tdks}) are integrated in real space in a cuboid box of $29\times29\times67$ a.u.~with a space discretization of 0.4 a.u.~employing a nine-point stencil for the spatial derivatives and a fourth-order Taylor approximation to the time-evolution operator. The time-dependent induced dipole moment is the first moment of the induced electron density $\Delta n(\mathbf r, t) = n(\mathbf r, t) - n(\mathbf r, -\infty)$. 
We choose a UV-VIS field $F_L(t)$ resembling those currently attainable \cite{graf_intense_2008,reiter_generation_2010} and record the three-dimensional microscopic electric field distribution $F_{\mathrm{NF}}(\mathbf r, t)$ evaluated as the negative gradient of the induced potential $V(\mathbf r, t) - V(\mathbf r, -\infty)$ (Eq.~(\ref{eq:tdks})) \cite{zuloaga_quantum_2009,*zuloaga_quantum_2010,stella_performance_2013,zhang_ab_2014}. 

\begin{figure}
\centerline{\epsfig{file=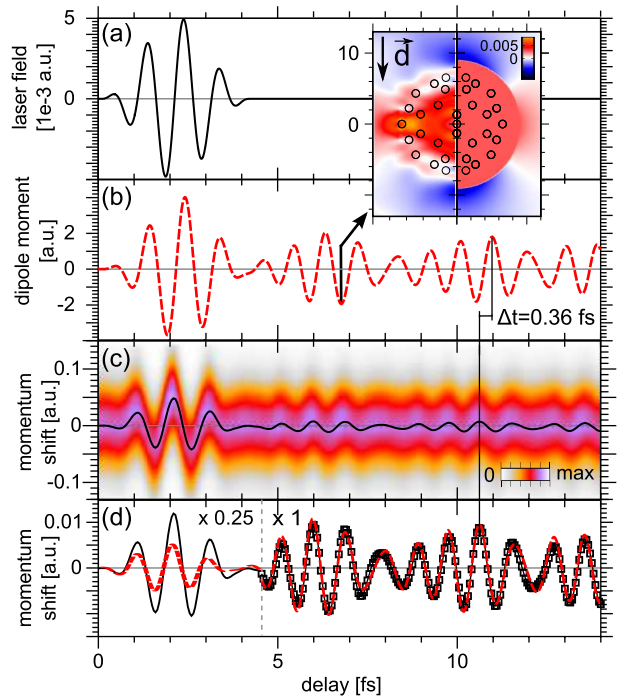,width=0.93 \columnwidth}}
%\centerline{\epsfig{file=fig1.eps,width=0.6 \columnwidth}}  %draft
\caption{(Color online) 
(a) Laser field (central wavelength 310 nm, 4 fs pulse length, $\sin^2$ envelope, intensity $9\xten{11}\Wcm$)
(b) Dipole moment of \Necsixty. 
Inset: Near-field (left) and near-field of dielectric sphere with the same dipole moment (right, Eq.~\ref{eq:field}, at $t=6.7$ fs). Circles mark the atomic positions.
(c) Simulated streaking spectrogram ($\omega_\mathrm{XUV}= 100 \eV$, duration 100 as). Black line: center of momentum streaking signal $\Delta P(\Delta \tau)$. 
(d) Enlarged streaking signal after the conclusion of the laser pulse (black solid, squares) and time shifted dipole moment (red dashed). 
}
\label{fig1}
\end{figure}

The ab-initio near-field distribution serves as input for subsequent classical trajectory Monte Carlo (CTMC) simulations \cite{lemell_classical-quantum_2013}. 
For the propagation of the photoemitted electron the CTMC simulation of streaking is well justified for attosecond XUV pulses with pulse duration $\tau_\mathrm{XUV}$ short compared to the oscillation period of the streaking field, $\tau_\mathrm{XUV} \ll T_d$ ($T_d$: period of the dipole streaking field)
when interference effects are negligible and for moderately fast electrons (velocity $v \gtrsim 1 \au$) \cite{nagele_time-resolved_2011,pazourek_attosecond_2015}.
The initial distribution of the photoelectrons is given by the neon 2s spatial distribution with an initial dipolar angular distribution with anisotropy parameter $\beta_\mathrm{2s} = 2$.
Their initial mean energy is given by $E_i = \hbar \omega_\mathrm{XUV} - I_p$ with 2s binding energy $I_p = 48.5 $ eV taking into account spectral broadening by the attosecond XUV pulse (duration $\tau_\mathrm{XUV}=100 \as$). The ionization probability is assumed proportional to the instantaneous XUV intensity. The electrons are propagated in the space and time dependent molecular near-field $F_\mathrm{NF}(\mathbf r,t)$. Their final momenta within an acceptance angle of 10 deg from the polarization direction are recorded as a function of the delay $\Delta \tau$ between XUV and UV-VIS pulses in a ``streaking spectrogram'' (Fig.~\ref{fig1}c). The streaking signal of each simulation is the delay-dependent shift of the center of momentum, $\Delta P(\Delta \tau)$. 

Elastic scattering at the cores of the carbon cage atoms and inelastic electron-electron scattering along the trajectory can by easily taken into account within the CTMC simulation \cite{solleder_spin-dependent_2007,lemell_simulation_2009,neppl_direct_2015} but do not significantly influence the near-field induced shift of the mean momentum. 
The most likely source for the uncertainty in the simulation comes from neglecting the back-action of the core hole on the response of \Necsixty. The time scale on which the \csixty~spectator electrons will respond to the additional Coulomb potential of the ionized neon core can be estimated from TD-DFT simulations to be around 10 a.u. (0.25 fs) \cite{muino_time-dependent_2011,koval_dynamic_2012} starting from the time the electron passes the \csixty~shell. The ionized electron has then already reached a distance of $\sim 20 \au$ or three times the \csixty~radius from the shell and has sampled most of the (de)acceleration exerted by the molecular near-field.

%Additionally, we neglect some reaction channels like elastic scattering from the carbon cores and inelastic losses from electron-electron collisions with the cage electrons. These may lead to additional broadening of the streaking spectrograms but will not impact the near-field induced shift of the center of momentum.

During the laser pulse, the dipole moment oscillates in phase with the electric field (Fig.~\ref{fig1}b) while the streaking shift $\Delta P$ is in phase with the vector potential (Fig.~\ref{fig1}c), i.e.~90 deg shifted relative to the field as in conventional streaking. 
After the conclusion of the laser pulse, the time-dependent dipole moment continues to oscillate (Fig.~\ref{fig1}b) with a dominant oscillation period $\sim 1 \fs$ corresponding  to the transition frequency to the lowest dipole-allowed excited states $E_e - E_0 \sim 4 \eV$. The longer beating period $\sim 5 \fs$ corresponds to the energy difference between different coherently excited states $E_{e \prime} - E_e \sim 1 \eV$. This time dependence of the dipole moment indicates changes in the induced density as well as in the alternating dipolar near-field (Fig.~\ref{fig1}b inset). The nano-scale electric field distribution resembles the sum of the laser $\mathbf F_L(t) = F_L(t) \hat z$ and of the field of a dielectric sphere with dipole moment $\mathbf d(t) = d(t) \hat z$,  
\begin{equation} \label{eq:field}
  \mathbf F_\mathrm{NF}(\mathbf r, t) =  \mathbf F_L(t) + \left\{
  \begin{array}{lr}
     -\mathbf d(t)/R^3                                                   & \mathrm{for} \quad r < R   \\
     -\mathbf d(t)/r^3 + 3 \frac{\mathbf d(t) \cdot \mathbf r }{r^5} \mathbf r & \mathrm{for} \quad r > R
  \end{array}
  \right. \quad ,
\end{equation}
where $R=9$ a.u.~matches the field enhancement and near-field of the effective electronic surface somewhat outside the ionic positions at 6.66 a.u. (Fig. \ref{fig1}b inset) \cite{neppl_direct_2015}. 
For time delays exceeding the duration of the UV-VIS pulse, the streaking sprectrogram encodes the influence of the time-dependent dipolar near-field on the emitted electron.
We find that the streaking modulation $\Delta P(\Delta \tau)$ traces the oscillations of the dipole moment to a surprisingly good degree of approximation. The molecular dipole moment $d(t)$ can be reconstructed from the streaking signal $\Delta P(\Delta \tau)$ as 
\begin{equation} \label{eq:reconstruct}
  d(\Delta \tau) = A \, \Delta P(\Delta \tau + \Delta t)
\end{equation}
with a scaling factor $A \simeq 190 \au$ and a time shift $\Delta t \simeq 0.36 \fs$ (Fig.~\ref{fig1}c). 

\begin{figure}
\centerline{\epsfig{file=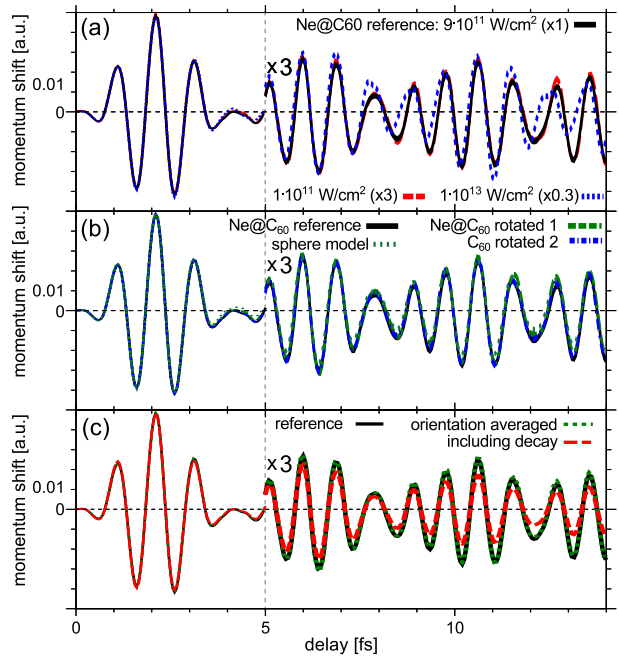,width= 0.93 \columnwidth}}
%\centerline{\epsfig{file=fig2.eps,width=0.6 \columnwidth}} %draft
\caption{(Color online) 
(a) Intensity dependence of streaking signal; reference calculation from Fig.~\ref{fig1}c ($9\xten{11} \Wcm$, black solid); lower intensity ($1\xten{11} \Wcm$, red dashed, $\times 3$), higher intensity ($1\xten{13} \Wcm$, blue dotted, $\times 0.3$) each scaled by streaking field amplitude. Streaking momentum shift is 0.007 a.u.~(0.17 eV), 0.19 a.u.~(0.51 eV), 0.67 a.u.~(1.82 eV) respectively. 
(b) Dependence on near-field; reference calculation from Fig.~\ref{fig1}c (black solid), different orientation of \Necsixty~(green dashed) and of \csixty (blue dash-dotted), sphere model (Eq.~\ref{eq:field}, turquoise dotted). %, analytical model for interval $5-7$ fs (Eq.~\ref{eq:ana}, violet long dashed). 
(c) Decay of dipole moment; reference calculation from Fig.~\ref{fig1}c (black solid) for one orientation, orientation averaged (green dotted), orientation average with exponential decay with decoherence time $t_c = 20 \fs$ (details see text, red dashed). 
}
\label{fig2}
\end{figure}

The mapping of the molecular dipole moment onto the streaking signal (Eq.~\ref{eq:reconstruct}) is remarkably robust under parameter variation. 
We numerically obtain a linear dependence of the streaking signal on the pump laser field amplitude while its temporal shape remains almost unchanged when varying the intensity over 2 orders of magnitude (Fig.~\ref{fig2}a). % for intensities up to $1\xten{13} \Wcm$ . 
%At this intensity, the  momentum shift amplitude reaches 0.67 a.u.~(or $1.82 \eV$). 
Similarly, the streaking time signal does not depend sensitively on details of the near-field. When either changing the molecular orientation, or removing the central neon atom, or even replacing the field distribution from TD-DFT by the polarized sphere Eq.~\ref{eq:field} (Fig.~\ref{fig2}b), the mapping of the time-dependent induced dipole moment to the streaking signal closely agrees with that of the reference calculation (Fig.~\ref{fig1}c). 
Averaging over molecular orientations and, thus, over the anisotropy of the molecular polarizability, we find that the orientation-induced dephasing is negligible on the time scale of our simulation. % with a spread of about $0.01 \fs$. 

\begin{figure}
\centerline{\epsfig{file=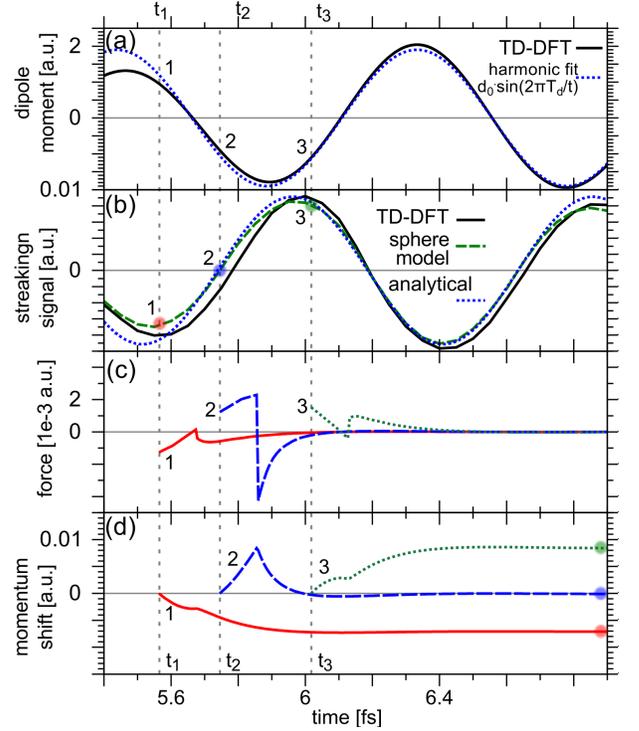,width=0.93 \columnwidth}}
%\centerline{\epsfig{file=fig3.eps,width=0.6 \columnwidth}} %draft
\caption{(Color online) 
Enlarged time interval $[5.4\fs-6.9\fs]$ from Fig.~\ref{fig2}. 
(a) Dipole moment; reference TD-DFT calculation from Fig.~\ref{fig1} (black solid), single-harmonic fit (blue dotted). 
(b) Streaking signal; reference TD-DFT calculation from Fig.~\ref{fig1} (black solid), simplified sphere field distribution (Eq.~\ref{eq:field}, green dashed), analytical model (Eq.~\ref{eq:ana}, blue dotted). 
(c) Force along electron trajectories starting at $t_1 = 5.56$ fs, $t_2 = 5.74$ fs, $t_3 = 6.02$ fs (sphere field distribution Eq.~\ref{eq:field}). 
%$t_1 = 5.56$ fs, $t_2 = 5.96$ fs, $t_3 = 6.02$ fs (sphere model). 
(d) Momentum shift accumulated along these trajectories. 
\label{fig3}
}
\end{figure}

We now analyze the properties underlying the remarkably stable mapping of the molecular dipole moment onto the streaking signal $\Delta P$, employing the simplified field distribution (Eq.~\ref{eq:field}) for convenience. 
The field distribution in the absence of the laser pulse (Eq.~\ref{eq:field} with laser $F_L(t) = 0$ and $d(t) > 0$) is symmetric under reflection ($z \to -z$) at a plane perpendicular to the dipole polarization. Correspondingly, the electrostatic potential is antisymmetric and is zero both at the origin and at infinity. 
An electron emitted from the center of the \csixty~will emerge at large distances with its initial energy as the deceleration inside the sphere is exactly canceled by the acceleration in the enhanced near-field outside for a static field distribution (Eq.~\ref{eq:field}). 
The observed delay time dependent momentum shift $\Delta P$ and, thus, the self-streaking relies on non-adiabatic corrections to the static field. 
The time scale for the traversal of the molecular near-field extending over $\simeq 2 R$ for a photoelectron with velocity $v_0 \simeq \sqrt{ 2 (\hbar \omega_\mathrm{XUV} - I_p)} \approx 2 \au$ is comparable to a fraction of the dipole oscillation period $\simeq T_d/4$, over which the field changes from zero to its maximum value, $2 R / v_0 \simeq T_d/4$. 
This qualitative argument can be made quantitative by further simplifying Eq.~\ref{eq:field}. 
Accordingly, the streaking momentum shift for an electron ionized at birth time $\Delta \tau$ along the laser polarization direction $z$ is approximately given by 
\begin{eqnarray} \label{eq:ana}
\Delta P(\Delta \tau) &=& \int_{\Delta \tau}^{\infty} F_\mathrm{NF}\left( z(t), t \right) \,\, dt \approx \nonumber \\
&\approx -& \frac{d_0 T_d}{R^3 \pi} \sin\left(\pi \frac{R}{v_0 T_d}\right) \sin\left(\pi \frac{R+2 v_0 \Delta \tau}{v_0 T_d}\right) \nonumber \\
& + & \frac{d_0}{R^2 T_d^2 v_0^3} \Bigg\{ 2 \pi R v_0 T_d \cos\left( 2 \pi \frac{ R + v_0 \Delta \tau}{v_0 T_d}\right)  + \nonumber \\
& & v_0^2 T_d^2 \sin\left( 2 \pi \frac{R + v_0 \Delta \tau}{v_0 T_d}\right) + \nonumber \\
& & 2 \pi^2 R^2 \Big[ 2 \mathrm{Ci}\left( \frac{2 \pi R}{ v_0 T_d} \right) \sin\left( 2 \pi \frac{ \Delta \tau}{ T_d} \right) - \nonumber \\
& & \, \left(\pi - 2 \mathrm{Si}\left(\frac{2 \pi R}{v_0 T_d}\right) \right) \cos\left(2 \pi \frac{\Delta \tau}{T_d}\right) \Big] \Bigg\} \,\,\, .
\end{eqnarray}
%%%%%%%%%
%    old version
%%%%%%%%%
%\begin{eqnarray} \label{eq:ana}
%\Delta P(t_b) &=& \int_{t_b}^{\infty} F_\mathrm{NF}\left( z(t), t \right) \,\, dt \approx \nonumber \\
%                     &\approx& \frac{d_0 T_d }{R^2} \Bigg\{  \frac{-1}{2 \pi R} \sin\left(\pi \frac{R}{v_0 T_d}\right) \sin\left(\pi \frac{R/v_0 + 2 t_b}{T_d}\right) \nonumber \\
%                      & & + \frac{R \pi}{R^2 \pi^2 + v_0^2 T_d^2} \bigg[ \cos\left(2 \pi \frac{t_b + R/v_0}{T_d}\right) \nonumber \\ 
%                     & & + \frac{T_d v_0}{\pi R} \sin\left(2 \pi \frac{t_b + R/v_0}{T_d}\right) \bigg] \Bigg\} \quad .
%\end{eqnarray}
%Here we have for simplicity assumed 
%(i) the dipole moment oscillates with a single Fourier component $d(t) = d_0 \sin(2 \pi t_b/T_d)$; 
%(ii) a straight-line trajectory $z(t) = v_0 (t-t_b)$ of the emitted electron; and 
%(iii) an exponential rather than power-law decay of the dipolar near-field with the same value at $R$ and integral to $\infty$ $2 d(t) / z^3 \to (2 e^2 d(t) / R^3) \exp(- 2z/R)$ to facilitate the analytic evaluation of the integral. 
%Eq.~\ref{eq:ana} explicitly displays the dependence on $T_d$, $v_0$, and $R$. 
%approximately reproduces the amplitude and phase of the full simulation (Fig.~\ref{fig3}), including the numerically observed linear scaling with the oscillation amplitude, and encodes the sensitive dependence on $R$, $v_0$ and $T_d$. 
Here we have assumed for simplicity 
(i) the dipole moment oscillates with a single Fourier component $d(t) = d_0 \sin(2 \pi t/T_d)$ and 
(ii) a straight-line trajectory $z(t) = v_0 (t-\Delta \tau)$ of the emitted electron. 
The first term represents the motion inside the cage and the second term the motion in the enhanced near-field. 
%The cosine and sine integral functions may be approximated by their series expansions $\mathrm{Ci}(x) \sim \gamma_E + \log(x) - x^2/4 + \ldots \approx 0.18$ with Euler's constant $\gamma_E \approx 0.58$ and $\mathrm{Si}(x) \sim x - x^3/18 + \ldots \approx 0.76$. 
Eq.~\ref{eq:ana} effectively predicts values for $A$ and $\Delta t$ (Eq.~\ref{eq:reconstruct}) in good agreement with the simulation (see Fig.~\ref{fig3}b). 
The largest momentum shift is incurred for electrons emitted shortly after the extrema of the dipole moment  (Fig.~\ref{fig3}, times $t_1$ and $t_3$). These trajectories are accelerated inside the cage and reach the molecular surface at the time when the dipole moment changes its sign so that they are further accelerated by the enhanced near-field. For electrons born at a time when the magnitude of the instantaneous dipole moment increases (Fig.~\ref{fig3}, $t_2$), the self-streaking field is stronger but the field inside and outside the cage have different signs and therefore partially compensate each other leading to a reduced momentum shift. %The maximum of the streaking signal lags behind the maximum of the dipole moment by about $0.3 T_d$. 
The present self-streaking protocol is independent of how the dipole excitation is initiated. Alternatively to the moderately intense VIS-UV pulses employed here, lower frequency pulses of higher intensity could be used as well \cite{li_coherent_2015}. % optical range  \cite{rubio_collective_1994} 

%Eventually, the excess energy in the electronic system will dissipate either radiatively (typically on the nanosecond time scale) or diffuse into other degrees of freedom (transverse electronic or phononic).
Eventually, the coherent electronic dipole excitation will dephase by coupling to vibrational degrees and decay by electron-electron scattering or radiative relaxation, the latter of which occurring on the nanosecond scale. 
For \csixty, the fastest dephasing and dissipative process is estimated to be coupling to the lattice degrees of freedom. An estimate for its time scale is given by the oscillation period of the C$=$C stretching mode of $\sim 23 \fs$. This agrees well with estimates of electron-phonon coupling from the experimental spectra and theory \cite{leach_electronic_1992,gunnarsson_photoemission_1995,iwahara_vibronic_2010,faber_electron-phonon_2011,falke_coherent_2014}. 
Coupling to other degrees of freedom will lead to damping of the dipole oscillation amplitude and, in turn, of the streaking amplitude (Fig.~\ref{fig2}c). Self-streaking, thus, affords the opportunity to study dephasing and dissipative processes directly in the time domain and benchmark emerging ab-inito descriptions of non-adiabatic electron-ion coupling \cite{fischer_nonadiabatic_2014} and electron-electron scattering processes \cite{wijewardane_time-dependent_2005,dagosta_relaxation_2006}. 

In conclusion, we have demonstrated that dipolar excitations in the \Necsixty~molecule can be investigated with attosecond resolution employing a novel variant of the attosecond streaking protocol involving intrinsic rather than external oscillating strong fields. This ``self-streaking'' maps out the real-time modulations and decay of the time-dependent molecular near-field and dipole moment. For endohedral fullerenes, photoionization of the central rare-gas atom provides a localized source of distinguishable fast streaking electrons. Similarly, for studying processes in more complex systems such as organic molecules containing several atomic species, contributions from different locations within the molecule can be disentangled by the final electron energy. We may envision time-domain visualization of charge-transfer processes \cite{kuleff_electron-correlation-driven_2013,falke_coherent_2014} and hot electron dynamics in small clusters and nanoparticles \cite{schlipper_multiple_1998,klamroth_ultrafast_2009,nest_origin_2010} with sub-femtosecond resolution. 

%\section*{Acknowledgments}
This work was supported by the FWF (Austria), SFB-041 ViCoM, SFB-049 Next Lite, doctoral college W1243, and P21141-N16. G.W.\ thanks the IMPRS-APS for financial support. X.-M.T.\ was supported by a Grant-in-Aid for Scientific Research (C24540421) from the JSPS. K.Y.\ acknowledges support by the Grants-in-Aid for Scientific Research Nos.\ 23340113 and 25104702. Calculations were performed using the Vienna Scientific Cluster (VSC) and the supercomputer at the Institute of Solid State Physics, University of Tokyo.

%\vspace{-0.3cm}
\section*{References}
%\vspace{-1.0cm}

\bibliography{gw}   %USE : awk -f fix-bib.awk georgwachter.bib > gw.bib

%%% troubles: 
% there is a "stupid latex error" somewhere in the prl style when also using babel/bibtex.
% the quick solution is to delete the offending "language = {English}" line!
%  awk -f fix-bib.awk georgwachter.bib > gw.bib
%
% of course, this has to be repeated after every zotero export ... 
%%%

%%%% 
%%%% journal abbreviations and general improvements: 
%%%% -- run pdflatex / bibtex / ... 
%%%% -- cp dipolestreak.bbl dipolestreak.bbl.bak
%%%% -- awk -f change_bbl.awk dipolestreak.bbl.bak > dipolestreak.bbl
%%%% -- run pdflatex (BUT NOT BIBTEX) [ctrl-alt-p]
%}
% --- Comment Dani  ---
%while exporting from Zotero use encoding Western (ISO-8859-1) as bibtex cannot 
%handle UTF-8. Also use Format BibTeX instead of BibLaTeX.
%--- Comment End Dani --- 

% --- some snippets ---
%\tableofcontents
%\vspace{1cm}
%\section*{Introduction}
%---
%\begin{figure}
%\centerline{\epsfig{file=fig4.eps,width=7.5cm}}
%\caption{(Color online) \GW{caption} }
%\label{fig4}
%\end{figure}
%---
%\begin{thebibliography}{30}
%\bibitem{Petek1997} H.~Petek and S.~Ogawa, Prog. Surf. Sci. {\bf 56}, 239 (1997).
%\end{thebibliography}
%---
\end{document}